\documentclass[ aip, jmp, preprint,]{revtex4-1}
\usepackage{amsmath}
\usepackage{graphicx}
\usepackage{dcolumn}
\usepackage{bm}

\begin{document}

\preprint{}
\title[]{Identities for density functionals linking functionals of different
densities.}
\author{Daniel P. Joubert}
\email{daniel.joubert2@wits.ac.za}
\affiliation{Centre for Theoretical Physics, University of the Witwatersrand, PO Wits
2050, Johannesburg, South Africa}
\date{\today }

\begin{abstract}
In electron density functional theory formal properties of density
functionals play an important role in constructing and testing approximate
functionals. In this paper it is shown that the equation
\begin{eqnarray*}
&&G^{\gamma }\left[ \rho _{N}\right] -G^{\gamma }\left[ \rho _{N-1}^{\gamma }%
\right] \\
&=&\int d^{3}r\left( \rho _{N}\left( \mathbf{r}\right) -\rho _{N-1}^{\gamma
}\left( \mathbf{r}\right) \right) \frac{\delta G^{\gamma }\left[ \rho _{N}%
\right] }{\delta \rho _{N}\left( \mathbf{r}\right) }
\end{eqnarray*}%
is satisfied by a number of functionals that are of interest in density
functional theory. In particular it is satisfied by $E_{hxc}^{\gamma }\left[
\rho \right] ,$ the sum of the Hartree and exchange-correlation energies.
The parameter $\gamma $ scales the mutual electron Coulomb interaction
energy while $\rho _{N}\left( \mathbf{r}\right) $ and $\rho _{N-1}^{\gamma
}\left( \mathbf{r}\right) $ are $N$-electron and $(N-1)$-electron densities
determined from the same adiabatic scaled external potential of the $N$%
-electron system at coupling strength $\gamma $.
\end{abstract}

\pacs{31.15.E-,71.15.Mb,71.10.-M,71.45.Gm}
\keywords{Density functional theory, Electronic structure theory, Exchange
and correlation energy, Kinetic energy, Coulomb interaction energy, Atoms,
molecules, and solids }
\maketitle














\section{ Introduction}

The Kohn-Sham (KS) formulation \cite{KohnSham:65} of Density Functional
Theory (DFT) \cite{HohenbergKohn:64} is one of the most important tools for
the calculation of electronic structure of molecules and solids. In all
practical applications of DFT, however, approximations to the exact
functionals have to be made \cite%
{AdriennRuzsinszky2011,JohnP.Perdew2009,GaborI.Csonka2009a,Staroverov2004,TaoPerdew03}
Exact relations for density functionals and density functional derivatives
can play an important role in the development of accurate approximations to
the exact functionals. A successful approach to the design of improved
approximate density functionals is by 'constraint satisfaction'\cite%
{JohnP.Perdew2005}, where the approximate functionals are required to
satisfy properties of the exact functionals. With this in mind, the
following relations for DFT energy functionals are derived:%
\begin{eqnarray}
&&F^{\gamma }[\rho _{N}]-F^{\gamma }[\rho _{N-1}^{\gamma }]  \notag \\
&=&\int d^{3}r\left( \rho _{N}\left( \mathbf{r}\right) -\rho _{N-1}^{\gamma
}\left( \mathbf{r}\right) \right) \frac{\delta F^{\gamma }\left[ \rho _{N}%
\right] }{\delta \rho _{N}\left( \mathbf{r}\right) }  \label{id1}
\end{eqnarray}%
\begin{eqnarray}
&&V_{ee}^{\gamma }\left[ \rho _{N}\right] -V_{ee}^{\gamma }\left[ \rho
_{N-1}^{\gamma }\right]  \notag \\
&=&\int d^{3}r\left( \rho _{N}\left( \mathbf{r}\right) -\rho _{N-1}^{\gamma
}\left( \mathbf{r}\right) \right) \frac{\delta V_{ee}^{\gamma }\left[ \rho
_{N}^{1}\right] }{\delta \rho _{N}^{1}\left( \mathbf{r}\right) }  \label{id3}
\end{eqnarray}%
\begin{eqnarray}
&&T^{\gamma }\left[ \rho _{N}\right] -T^{\gamma }\left[ \rho _{N-1}^{\gamma }%
\right]  \notag \\
&=&\int d^{3}r\left( \rho _{N}\left( \mathbf{r}\right) -\rho _{N-1}^{\gamma
}\left( \mathbf{r}\right) \right) \frac{\delta T^{\gamma }\left[ \rho _{N}%
\right] }{\delta \rho _{N}\left( \mathbf{r}\right) }  \label{id2}
\end{eqnarray}%
\begin{eqnarray}
&&T_{c}^{\gamma }\left[ \rho _{N}\right] -T_{c}^{\gamma }\left[ \rho
_{N-1}^{\gamma }\right]  \notag \\
&=&\int d^{3}r^{\prime }\left( \rho _{N}\left( \mathbf{r}^{\prime }\right)
-\rho _{N-1}^{\gamma }\left( \mathbf{r}^{\prime }\right) \right) \frac{%
\delta T_{c}^{\gamma }\left[ \rho _{N}\right] }{\delta \rho _{N}\left(
\mathbf{r}\right) }.  \label{id4}
\end{eqnarray}%
\begin{eqnarray}
&&T^{0}\left[ \rho _{N}\right] -T^{0}\left[ \rho _{N-1}^{\gamma }\right]
\notag \\
&=&\int d^{3}r^{\prime }\left( \rho _{N}\left( \mathbf{r}^{\prime }\right)
-\rho _{N-1}^{\gamma }\left( \mathbf{r}^{\prime }\right) \right) \frac{%
\delta T^{0}\left[ \rho _{N}\right] }{\delta \rho _{N}\left( \mathbf{r}%
\right) }.  \label{id5}
\end{eqnarray}%
\begin{eqnarray}
&&E_{hxc}^{\gamma }\left[ \rho _{N}\right] -E_{hxc}^{\gamma }\left[ \rho
_{N-1}^{\gamma }\right]  \notag \\
&=&\int d^{3}r^{\prime }\left( \rho _{N}\left( \mathbf{r}^{\prime }\right)
-\rho _{N-1}^{\gamma }\left( \mathbf{r}^{\prime }\right) \right) \frac{%
\delta E_{hxc}^{\gamma }\left[ \rho _{N}\right] }{\delta \rho _{N}\left(
\mathbf{r}\right) }.  \label{id6}
\end{eqnarray}

Here $\rho _{N}$ and $\rho _{N-1}^{\gamma }$ are the ground state charge
densities of an interacting $N$ and $\left( N-1\right) $ electron system
with the same single particle multiplicative external potential $v_{\text{ext%
}}^{\gamma }\left( [\rho _{N}]\right) $. The potential $v_{\text{ext}%
}^{\gamma }\left( [\rho _{N}]\right) $ is constructed to keep the charge
density of the $N$ electron system independent of the coupling strength
parameter $\gamma $ \cite%
{HarrisJones:74,LangrethPerdew:75,LangrethPerdew:77,GunnarsonLundqvist:76}
that scales the electron-electron interaction strength. At $\gamma =1$ full
strength Coulomb interaction between electrons is included and the external
potential $v_{\text{ext}}^{\gamma }\left( [\rho _{N}]\right) $ is the
external potential of the fully interacting system, while $\gamma =0$
corresponds to the non-interacting Kohn-Sham potential. $F^{\gamma }[\rho ]$
is the sum of the kinetic energy functional $T^{\gamma }\left[ \rho \right] $
and the mutual Coulomb interaction energy $\gamma V_{ee}^{\gamma }\left[
\rho \right] $ at interaction strength $\gamma $ and density $\rho .$ $%
T_{c}^{\gamma }\left[ \rho \right] $ is the correlation part of the kinetic
energy and $E_{hxc}^{\gamma }\left[ \rho \right] $ is the sum of the Hartree
and exchange-correlation energy.

As corollaries recursion relations are satisfied. For $G^{\gamma }\left[
\rho \right] =F^{\gamma }[\rho ],$ $T^{\gamma }[\rho ],$ $V_{ee}^{\gamma }%
\left[ \rho \right] ,$ $T_{c}^{\gamma }\left[ \rho \right] ,$ $T^{0}\left[
\rho \right] $ or $E_{hxc}^{\gamma }\left[ \rho \right] ,$%
\begin{eqnarray}
&&G^{\gamma }\left[ \rho _{N}\right] -G^{\gamma }\left[ \rho _{N-M}^{\gamma }%
\right]  \notag \\
&=&\sum_{L=0}^{M-1}\int d^{3}r\left( \rho _{N-L}^{\gamma }\left( \mathbf{r}%
\right) -\rho _{N-1-L}^{\gamma }\left( \mathbf{r}\right) \right) \frac{%
\delta G^{\gamma }\left[ \rho _{N-L}^{\gamma }\right] }{\delta \rho
_{N-L}^{\gamma }\left( \mathbf{r}\right) },\qquad M\leq N-1.
\end{eqnarray}%
Here the densities $\rho _{N-L}^{\gamma }\left( \mathbf{r}\right) $ are the (%
$N-L$)-particle densities derived from the groundstate wavefunctions of ($%
N-L $)-particle Hamiltonians with mutual Coulomb interaction $\gamma \hat{V}%
_{ee} $ and external potential $v_{\text{ext}}^{\gamma }\left( [\rho
_{N}]\right) . $ For $G^{\gamma }\left[ \rho \right] =V_{ee}^{\gamma }\left[
\rho _{N}\right] ,$ $T_{c}^{\gamma }\left[ \rho _{N}\right] $ or $%
E_{hxc}^{\gamma }\left[ \rho _{N}\right] ,\ G^{\gamma }\left[ \rho
_{1}^{\gamma }\right] =0$ and the functionals can be written as

\begin{equation}
G^{\gamma }\left[ \rho _{N}\right] =\sum_{L=0}^{N-2}\int d^{3}r\left( \rho
_{N-L}^{\gamma }\left( \mathbf{r}\right) -\rho _{N-1-L}^{\gamma }\left(
\mathbf{r}\right) \right) \frac{\delta G^{\gamma }\left[ \rho _{N-L}^{\gamma
}\right] }{\delta \rho _{N-L}^{\gamma }\left( \mathbf{r}\right) }.
\end{equation}

These identities relate functionals at different particle numbers and are
likely to place stringent restrictions on the possible forms that potential
approximate functionals can assume.

\section{Proof of equations 1-3}

In the adiabatic connection approach \cite%
{HarrisJones:74,LangrethPerdew:75,LangrethPerdew:77,GunnarsonLundqvist:76}
of the constrained minimization formulation of density functional theory
\cite{HohenbergKohn:64,KohnSham:65,Levy:79,LevyPerdew:85} the Hamiltonian $%
\hat{H}^{\gamma }$ for a system of $N$ electrons is given by
\begin{equation}
\hat{H}^{\gamma }=\hat{T}+\gamma \hat{V}_{ee}+\hat{v}_{N,\text{ext}}^{\gamma
}\left[ \rho _{N}\right] .  \label{a3}
\end{equation}%
Atomic units, $\hbar =e=m=1$ are used throughout. $\hat{T}$ is the kinetic
energy operator,%
\begin{equation}
\hat{T}=-\frac{1}{2}\sum_{i=1}^{N}\nabla _{i}^{2},  \label{a4}
\end{equation}%
\ and $\gamma \hat{V}_{ee\text{ }}$is a scaled electron-electron interaction,%
\begin{equation}
\gamma \hat{V}_{ee}=\gamma \sum_{i<j}^{N}\frac{1}{\left\vert \mathbf{r}_{i}-%
\mathbf{r}_{j}\right\vert }.  \label{a2}
\end{equation}%
The the external potential
\begin{equation}
\hat{v}_{\text{ext}}^{\gamma }\left[ \rho _{N}\right] =\sum_{i=1}^{N}v_{%
\text{ext}}^{\gamma }\left( \left[ \rho _{N}\right] ;\mathbf{r}_{i}\right) ,
\label{a1}
\end{equation}%
is constructed to keep the charge density fixed at $\rho _{N}\left( \mathbf{r%
}\right)$, the ground state charge density of the fully interacting system ($%
\gamma =1$), for all values of the coupling constant $\gamma.$ The external
potential has the form \cite{LevyPerdew:85,GorlingLevy:93}
\begin{align}
v_{\text{ext}}^{\gamma }(\left[ \rho _{N}\right] ;\mathbf{r})& =\left(
1-\gamma \right) v_{hx}([\rho _{N}];\mathbf{r})  \notag \\
& +v_{c}^{1}([\rho _{N}];\mathbf{r)}-v_{c}^{\gamma }([\rho _{N}];\mathbf{r)+}%
v_{\text{ext}}^{1}(\left[ \rho _{N}\right] ;\mathbf{r}),  \label{e1}
\end{align}%
where $v_{\text{ext}}^{1}(\left[ \rho _{N}\right] ;\mathbf{r})=v_{\text{ext}%
}\left( \mathbf{r}\right) $ is the external potential at full coupling
strength, $\gamma =1,$ and $v_{\text{ext}}^{0}(\left[ \rho _{N}\right] ;%
\mathbf{r})$ is the non-interacting Kohn-Sham potential. The exchange plus
Hartree potential \cite{ParrYang:bk89,DreizlerGross:bk90} $v_{hx}([\rho
_{N}];\mathbf{r}),$ is independent of $\gamma ,$ while the correlation
potential $v_{c}^{\gamma }([\rho _{N}];\mathbf{r)}$ depends in the scaling
parameter $\gamma .$

The chemical potential%
\begin{equation}
\mu =E_{N}^{\gamma }\left( v_{\text{ext}}^{\gamma }\left[ \rho \right]
\right) -E_{N-1}^{\gamma }\left( v_{\text{ext}}^{\gamma }\left[ \rho \right]
\right)  \label{e2}
\end{equation}%
depends on the asymptotic decay of the charge density \cite%
{ParrYang:bk89,DreizlerGross:bk90,JonesGunnarsson:89}, and hence is
independent of the coupling constant $\gamma $ \cite%
{LevyGorling:96,LevyGorlingb:96}. In Eq. (\ref{e2}) $E_{N-1}^{\gamma }$ is
the groundstate energy of the $\left( N-1\right) $-electron system with the
same single-particle external potential $v_{\text{ext}}^{\gamma }\left( %
\left[ \rho _{N}\right] ;\mathbf{r}\right) $ as the $N$-electron system:%
\begin{eqnarray}
\hat{H}^{\gamma }\left\vert \Psi _{\rho _{M}^{\gamma }}^{\gamma
}\right\rangle &=&E_{M}^{\gamma }\left\vert \Psi _{\rho _{M}^{\gamma
}}^{\gamma }\right\rangle  \notag \\
\hat{H}^{\gamma } &=&\hat{T}+\gamma \hat{V}_{ee}+\hat{v}_{M,\text{ext}%
}^{\gamma }\left[ \rho _{N}\right]  \notag \\
\hat{v}_{M,\text{ext}}^{\gamma }\left[ \rho _{N}\right] &=&\sum_{i=1}^{M}v_{%
\text{ext}}^{\gamma }\left( \left[ \rho _{N}\right] ;\mathbf{r}_{i}\right)
\label{h1}
\end{eqnarray}%
Note that by construction of $v_{\text{ext}}^{\gamma }\left( \left[ \rho _{N}%
\right] ;\mathbf{r}\right) ,$ Eq. (\ref{e1}), $\rho _{N}=\rho _{N}^{1}$ is
independent of $\gamma ,$ but the groundstate density of the $\left(
N-1\right) $-electron system derived from the same external potential at
coupling strength $\gamma ,$ $\rho _{N-1}^{\gamma }$, will in general be a
function of $\gamma .$

Define the energy functional \cite{Levy:79,ParrYang:bk89,DreizlerGross:bk90}
\begin{eqnarray}
F^{\gamma }[\rho ] &=&\left\langle \Psi _{\rho }^{\gamma }\left\vert \hat{T}%
+\gamma \hat{V}_{ee}\right\vert \Psi _{\rho }^{\gamma }\right\rangle  \notag
\\
&=&\min_{\Psi \rightarrow \rho }\left\langle \Psi \left\vert \hat{T}+\gamma
\hat{V}_{ee}\right\vert \Psi \right\rangle .  \label{b1}
\end{eqnarray}%
According to the Levy constrained minimization formulation\cite{Levy:79},
the wavefunction $\left\vert \Psi _{\rho }^{\gamma }\right\rangle $ yields
the density $\rho $ and minimizes $\left\langle \Psi \left\vert \hat{T}%
+\gamma \hat{V}_{ee}\right\vert \Psi \right\rangle .$ $F^{\gamma }[\rho ]$
can be decomposed as \cite{ParrYang:bk89,DreizlerGross:bk90}%
\begin{equation}
F^{\gamma }[\rho ]=T^{0}\left[ \rho \right] +\gamma E_{hx}\left[ \rho \right]
+E_{c}^{\gamma }\left[ \rho \right] ,  \label{b2}
\end{equation}%
The correlation energy $E_{c}^{\gamma }\left[ \rho \right] $ is defined as%
\cite{LevyPerdew:85}
\begin{eqnarray}
E_{c}^{\gamma }\left[ \rho \right] &=&\left\langle \Psi _{\rho }^{\gamma
}\left\vert \hat{T}+\gamma \hat{V}_{ee}\right\vert \Psi _{\rho }^{\gamma
}\right\rangle  \notag \\
&&-\left\langle \Psi _{\rho }^{0}\left\vert \hat{T}+\gamma \hat{V}%
_{ee}\right\vert \Psi _{\rho }^{0}\right\rangle ,  \label{ec1}
\end{eqnarray}%
where $\left\vert \Psi _{\rho }^{0}\right\rangle $ is the Kohn-Sham
independent particle groundstate wavefunction that yields the same density
as the interacting system at coupling strength $\gamma .$ $E_{hx}\left[ \rho %
\right] $ is the sum of the Hartree and exchange energy%
\begin{equation}
E_{hx}\left[ \rho \right] =\left\langle \Psi _{\rho }^{0}\left\vert \hat{V}%
_{ee}\right\vert \Psi _{\rho }^{0}\right\rangle  \label{b3}
\end{equation}%
and the kinetic energy functional $T^{0}\left[ \rho \right] $ is given by.%
\begin{equation}
T^{0}\left[ \rho \right] =\left\langle \Psi _{\rho }^{0}\left\vert \hat{T}%
\right\vert \Psi _{\rho }^{0}\right\rangle .  \label{b4}
\end{equation}%
The full kinetic energy
\begin{eqnarray}
T^{\gamma }\left[ \rho \right] &=&\left\langle \Psi _{\rho }^{\gamma
}\left\vert \hat{T}\right\vert \Psi _{\rho }^{\gamma }\right\rangle  \notag
\\
&=&T^{0}\left[ \rho \right] +T_{c}^{\gamma }\left[ \rho \right] ,  \label{b5}
\end{eqnarray}%
with the correlation part of the kinetic energy defined as%
\begin{equation}
T_{c}^{\gamma }\left[ \rho \right] =\left\langle \Psi _{\rho }^{\gamma
}\left\vert \hat{T}\right\vert \Psi _{\rho }^{\gamma }\right\rangle
-\left\langle \Psi _{\rho }^{0}\left\vert \hat{T}\right\vert \Psi _{\rho
}^{0}\right\rangle .  \label{ec2}
\end{equation}%
Assuming that $F^{\gamma }[\rho ]$ is defined for non-integer electrons \cite%
{PPLB:82,ParrYang:bk89,DreizlerGross:bk90}, at the solution point%
\begin{equation}
\frac{\delta F^{\gamma }\left[ \rho _{N}\right] }{\delta \rho _{N}\left(
\mathbf{r}\right) }+v_{\text{ext}}^{\gamma }\left( \left[ \rho _{N}\right] ;%
\mathbf{r}\right) =\mu  \label{b6}
\end{equation}%
Note that by definition of $F^{\gamma }[\rho ]$%
\begin{eqnarray}
E_{N}^{\gamma }\left( v_{\text{ext}}^{\gamma }\left[ \rho _{N}\right]
\right) &=&F^{\gamma }[\rho _{N}]+\int d^{3}r\rho _{N}\left( \mathbf{r}%
\right) v_{\text{ext}}^{\gamma }\left( \left[ \rho _{N}\right] ;\mathbf{r}%
\right)  \notag \\
E_{N-1}^{\gamma }\left( v_{\text{ext}}^{\gamma }\left[ \rho _{N}\right]
\right) &=&F^{\gamma }[\rho _{N-1}^{\gamma }]+\int d^{3}r\rho _{N-1}^{\gamma
}\left( \mathbf{r}\right) v_{\text{ext}}^{\gamma }\left( \left[ \rho _{N}%
\right] ;\mathbf{r}\right) .  \label{b7}
\end{eqnarray}%
From (\ref{e2}) and (\ref{b7}),%
\begin{equation}
F^{\gamma }[\rho _{N}]-F^{\gamma }[\rho _{N-1}^{\gamma }]=\mu -\int
d^{3}r\left( \rho _{N}\left( \mathbf{r}\right) -\rho _{N-1}^{\gamma }\left(
\mathbf{r}\right) \right) v_{\text{ext}}^{\gamma }\left( \left[ \rho _{N}%
\right] ;\mathbf{r}\right) .  \label{b8}
\end{equation}%
Since%
\begin{equation}
\int d^{3}r\left( \rho _{N}\left( \mathbf{r}\right) -\rho _{N-1}^{\gamma
}\left( \mathbf{r}\right) \right) =1,  \label{b9}
\end{equation}%
it follows from (\ref{b6}), (\ref{b8}) and (\ref{b9}) that%
\begin{equation}
F^{\gamma }[\rho _{N}]-F^{\gamma }[\rho _{N-1}^{\gamma }]=\int d^{3}r\left(
\rho _{N}\left( \mathbf{r}\right) -\rho _{N-1}^{\gamma }\left( \mathbf{r}%
\right) \right) \frac{\delta F^{\gamma }\left[ \rho _{N}\right] }{\delta
\rho _{N}\left( \mathbf{r}\right) }.  \label{b10}
\end{equation}

In a recent paper, starting from the virial theorem for the interacting
system \cite{Joubert2011a}, the author showed that%
\begin{eqnarray}
&&V_{ee}^{\gamma }\left[ \rho _{N}^{1}\right] -V_{ee}^{\gamma }\left[ \rho
_{N-1}^{\gamma }\right]  \notag \\
&=&\int d^{3}r\frac{\delta V_{ee}^{\gamma }\left[ \rho _{N}^{1}\right] }{%
\delta \rho _{N}^{1}\left( \mathbf{r}\right) }\left( \rho _{N}^{1}\left(
\mathbf{r}\right) -\rho _{N-1}^{\gamma }\left( \mathbf{r}\right) \right) .
\label{vee}
\end{eqnarray}%
Combining (\ref{b10}) and (\ref{vee}) yields%
\begin{eqnarray}
&&T^{\gamma }\left[ \rho _{N}\right] -T^{\gamma }\left[ \rho _{N-1}^{\gamma }%
\right]  \notag \\
&=&\int d^{3}r\frac{\delta T^{\gamma }\left[ \rho _{N}\right] }{\delta \rho
_{N}\left( \mathbf{r}\right) }\left( \rho _{N}\left( \mathbf{r}\right) -\rho
_{N-1}^{\gamma }\left( \mathbf{r}\right) \right) .  \label{r1}
\end{eqnarray}

\section{Proof of equations 4-6}

The mutual electron-electron repulsion energy functional can be expressed as

\begin{equation}
V_{ee}^{\gamma }\left[ \rho _{N}^{\gamma }\right] =E_{hx}\left[ \rho \right]
+\frac{E_{c}^{\gamma }\left[ \rho \right] -T_{c}^{\gamma }\left[ \rho \right]
}{\gamma },  \label{v1}
\end{equation}%
which follows from Eq. (\ref{ec1}) and the definitions (\ref{b2}) and (\ref%
{b5}) \cite{ParrYang:bk89,DreizlerGross:bk90}. Substitute Eq. (\ref{v1}) in
Eq. (\ref{vee}), take the derivative with respect to $\gamma $:%
\begin{eqnarray}
&&\frac{d}{d\gamma }\frac{1}{\gamma }\left( \left( E_{c}^{\gamma }\left[
\rho _{N}\right] -T_{c}^{\gamma }\left[ \rho _{N}\right] \right) -\left(
E_{c}^{\gamma }\left[ \rho _{N-1}^{\gamma }\right] -T_{c}^{\gamma }\left[
\rho _{N-1}^{\gamma }\right] \right) \right)   \notag \\
&&-\int d^{3}r\frac{\partial \rho _{N-1}^{\gamma }\left( \mathbf{r}\right) }{%
\partial \gamma }v_{hx}\left( \left[ \rho _{N-1}^{\gamma }\right] ;\mathbf{r}%
\right)   \notag \\
&=&-\frac{1}{\gamma }\int d^{3}r\frac{\partial \rho _{N-1}^{\gamma }\left(
\mathbf{r}\right) }{\partial \gamma }\left( \gamma v_{hx}\left( \left[ \rho
_{N}\right] ;\mathbf{r}\right) +v_{c}\left( \left[ \rho _{N}\right] ;\mathbf{%
r}\right) \right)   \notag \\
&&+\frac{1}{\gamma }\int d^{3}r\frac{\partial \rho _{N-1}^{\gamma }\left(
\mathbf{r}\right) }{\partial \gamma }\frac{\delta T_{c}^{\gamma }\left[ \rho
_{N}\right] }{\delta \rho _{N}\left( \mathbf{r}\right) }  \notag \\
&&+\int d^{3}r^{\prime }\left( \rho _{N}\left( \mathbf{r}^{\prime }\right)
-\rho _{N-1}^{\gamma }\left( \mathbf{r}^{\prime }\right) \right) \frac{d}{%
d\gamma }\frac{1}{\gamma }\frac{\delta }{\delta \rho _{N}\left( \mathbf{r}%
\right) }\left( E_{c}^{\gamma }\left[ \rho _{N}\right] -T_{c}^{\gamma }\left[
\rho _{N}\right] \right) .  \label{v2}
\end{eqnarray}%
It follows from the definition of $E_{c}^{\gamma }[\rho _{N}],$ Eq. (\ref%
{ec1}) that \cite{LevyPerdew:85}
\begin{equation}
\frac{\partial }{\partial \gamma }E_{c}^{\gamma }[\rho _{N}]=\frac{1}{\gamma
}\left( E_{c}^{\gamma }[\rho _{N}]-T_{c}^{\gamma }[\rho _{N}]\right) .
\label{b13}
\end{equation}%
With the aid of Eqs. (\ref{v2}), (\ref{apf}) and (\ref{b13}), it can be
shown that%
\begin{eqnarray}
&&\frac{d}{d\gamma }\left( T_{c}^{\gamma }\left[ \rho _{N}\right] \right)
-T_{c}^{\gamma }\left[ \rho _{N-1}^{\gamma }\right]   \notag \\
&=&\frac{d}{d\gamma }\int d^{3}r^{\prime }\left( \rho _{N}\left( \mathbf{r}%
^{\prime }\right) -\rho _{N-1}^{\gamma }\left( \mathbf{r}^{\prime }\right)
\right) \frac{\delta T_{c}^{\gamma }\left[ \rho _{N}\right] }{\delta \rho
_{N}\left( \mathbf{r}\right) }.  \label{v6}
\end{eqnarray}%
Integrating with respect to $\gamma $ and taking into account that $T_{c}^{0}%
\left[ \rho \right] =0$ (see Eq. (\ref{ec2})), it follows that

\begin{eqnarray}
&&T_{c}^{\gamma }\left[ \rho _{N}\right] -T_{c}^{\gamma }\left[ \rho
_{N-1}^{\gamma }\right]  \notag \\
&=&\int d^{3}r^{\prime }\left( \rho _{N}\left( \mathbf{r}^{\prime }\right)
-\rho _{N-1}^{\gamma }\left( \mathbf{r}^{\prime }\right) \right) \frac{%
\delta T_{c}^{\gamma }\left[ \rho _{N}\right] }{\delta \rho _{N}\left(
\mathbf{r}\right) }.  \label{v7}
\end{eqnarray}%
As a consequence, from Eqs. (\ref{r1}) and (\ref{v7})
\begin{eqnarray}
&&T^{0}\left[ \rho _{N}\right] -T^{0}\left[ \rho _{N-1}^{\gamma }\right]
\notag \\
&=&\int d^{3}r^{\prime }\left( \rho _{N}\left( \mathbf{r}^{\prime }\right)
-\rho _{N-1}^{\gamma }\left( \mathbf{r}^{\prime }\right) \right) \frac{%
\delta T^{0}\left[ \rho _{N}\right] }{\delta \rho _{N}\left( \mathbf{r}%
\right) }.  \label{v8}
\end{eqnarray}

With
\begin{equation}
E_{hxc}^{\gamma }\left[ \rho \right] =\gamma E_{hx}\left[ \rho \right]
+E_{c}^{\gamma }\left[ \rho \right]  \label{v9}
\end{equation}%
it follows from Eqs. (\ref{vee}), (\ref{v1}) and (\ref{v7}) that%
\begin{eqnarray}
&&E_{hxc}^{\gamma }\left[ \rho _{N}\right] -E_{hxc}^{\gamma }\left[ \rho
_{N-1}^{\gamma }\right]  \notag \\
&=&\int d^{3}r^{\prime }\left( \rho _{N}\left( \mathbf{r}^{\prime }\right)
-\rho _{N-1}^{\gamma }\left( \mathbf{r}^{\prime }\right) \right) \frac{%
\delta }{\delta \rho _{N}\left( \mathbf{r}\right) }E_{hxc}^{\gamma }\left[
\rho _{N}\right] .
\end{eqnarray}

\section{Proof of recursion relations, equation 7}

The identities in Eqs. (\ref{id1}) to (\ref{id6}) all have the form

\begin{eqnarray}
&&G^{\gamma }\left[ \rho _{N}\right] -G^{\gamma }\left[ \rho _{N-1}^{\gamma }%
\right]  \notag \\
&=&\int d^{3}r^{\prime }\left( \rho _{N}\left( \mathbf{r}^{\prime }\right)
-\rho _{N-1}^{\gamma }\left( \mathbf{r}^{\prime }\right) \right) \frac{%
\delta }{\delta \rho _{N}\left( \mathbf{r}\right) }G^{\gamma }\left[ \rho
_{N}\right] .  \label{v12}
\end{eqnarray}%
These identities were derived with $\gamma =1,$ full interaction strength,
as the starting point. The arguments used before are equally applicable when
the Coulomb operator $\gamma \hat{V}_{ee}$ is replaced by $\beta \gamma \hat{%
V}_{ee}$, the coupling strength is scaled by $\beta $ and the external
potential at full coupling strength $\chi =1,$ is taken as $v_{\text{ext}%
}^{\gamma }\left( \left[ \rho _{N}\right] ;\mathbf{r}_{i}\right) .$ Here $%
\beta $ plays the role that $\gamma $ played before$.$ Recursion relations
can now be derived at coupling strength $\gamma $. If the single particle
external potential is kept fixed, i.e. for an $M$-electron system
\begin{equation}
\hat{v}_{\text{ext}}^{M,\gamma }\left[ \rho _{N}\right] =\sum_{i=1}^{M}v_{%
\text{ext}}^{\gamma }\left( \left[ \rho _{N}\right] ;\mathbf{r}_{i}\right)
\end{equation}%
and $\rho _{M}^{\beta =1}\equiv \rho _{M}^{\gamma }\left( v_{\text{ext}%
}^{\gamma }\left[ \rho _{N}\right] \right) $ is the $M$-electron density
constructed from a groundstate of the $M$-electron Hamiltonian $\hat{H}%
_{M}^{\beta =1}=\hat{T}+\gamma \hat{V}_{ee}+\hat{v}_{\text{ext}}^{M,\gamma }%
\left[ \rho _{N}\right] .$ Thus, from Eq. (\ref{v12}), but starting from $%
\rho _{N-1}^{\gamma },$
\begin{eqnarray}
&&G^{\gamma }\left[ \rho _{N-1}^{\gamma }\right] -G^{\gamma }\left[ \rho
_{N-2}^{\gamma }\right]  \notag \\
&=&\int d^{3}r\left( \rho _{N-1}^{\gamma }\left( \mathbf{r}\right) -\rho
_{N-2}^{\gamma }\left( \mathbf{r}\right) \right) \frac{\delta G^{\gamma }%
\left[ \rho _{N-1}^{\gamma }\right] }{\delta \rho _{N-1}^{\gamma }\left(
\mathbf{r}\right) }.  \label{c2}
\end{eqnarray}%
Continuing this pattern leads to
\begin{eqnarray}
&&G^{\gamma }\left[ \rho _{N}\right] -G^{\gamma }\left[ \rho _{N-M}^{\gamma }%
\right]  \notag \\
&=&\sum_{L=0}^{M-1}\int d^{3}r\left( \rho _{N-L}^{\gamma }\left( \mathbf{r}%
\right) -\rho _{N-1-L}^{\gamma }\left( \mathbf{r}\right) \right) \frac{%
\delta G^{\gamma }\left[ \rho _{N-L}^{\gamma }\right] }{\delta \rho
_{N-L}^{\gamma }\left( \mathbf{r}\right) },\qquad M\leq N-1.  \label{c3}
\end{eqnarray}%
For $G^{\gamma }\left[ \rho _{N}\right] =V_{ee}^{\gamma }\left[ \rho _{N}%
\right] ,$ $T_{c}^{\gamma }\left[ \rho _{N}\right] $ or $E_{hxc}^{\gamma }%
\left[ \rho _{N}\right] ,G^{\gamma }\left[ \rho _{1}^{\gamma }\right] =0,$
hence for $V_{ee}^{\gamma }\left[ \rho _{N}\right] ,T_{c}^{\gamma }\left[
\rho _{N}\right] $ and $E_{hxc}^{\gamma }\left[ \rho _{N}\right] $

\begin{equation}
G^{\gamma }\left[ \rho _{N}\right] =\sum_{L=0}^{N-2}\int d^{3}r\left( \rho
_{N-L}^{\gamma }\left( \mathbf{r}\right) -\rho _{N-1-L}^{\gamma }\left(
\mathbf{r}\right) \right) \frac{\delta G^{\gamma }\left[ \rho _{N-L}\right]
}{\delta \rho _{N-L}^{\gamma }\left( \mathbf{r}\right) }.  \label{c4}
\end{equation}

\section{Discussion and summary}

The identities reported here were derived by reference to the eigenfunctions
of many particle Hamiltonians, though indirectly, via Eq. (\ref{b6}). They
therefore are valid for $w$-representable densities \cite%
{ParrYang:bk89,DreizlerGross:bk90}, in other words for densities that can be
determined from groundstate wavefunctions of a many-particle Hamiltonian.
The assumption was made that all functional derivatives are well behaved and
this implies that the functionals are defined for non-integer particle
numbers \cite{PPLB:82}. In the derivations no explicitly reference was made
to the structure of the wavefunctions or density, the equations are
therefore valid for all groundstate densities. Specifically they are valid
for all degenerate densities.

Equation (\ref{id6}) may be the most useful of the set presented here for
testing approximate exchange correlation functionals. The obvious test can
be done at full coupling strength, where for a real system the external
potential is the Coulomb interaction with the nuclei, and requires two
separate self-consistent Kohn-Sham calculations for the $N$ and $\left(
N-1\right) $ electron systems to determine $\rho _{N}^{1}\left( \mathbf{r}%
\right) $ and $\rho _{N-1}^{1}\left( \mathbf{r}\right) .$ For $T^{0}\left[
\rho \right] $ a single Kohn-Sham calculation is sufficient. The other
identities are more difficult to use for testing purposes since they are
satisfied by the functionals of the exact densities which will require
independent calculations to determine accurate densities.

If the functionals can be expanded as a series in the usual form\cite%
{ParrYang:bk89}
\begin{eqnarray}
&&G^{\gamma }\left[ \rho \right]  \notag \\
&=&G^{\gamma }\left[ \rho _{0}\right] +\sum_{n=1}^{\infty }\frac{1}{n!}\int
d^{3}r_{1}...d^{3}r_{n}\left( \rho \left( \mathbf{r}_{1}\right) -\rho
_{0}\left( \mathbf{r}_{1}\right) \right) ...\left( \rho \left( \mathbf{r}%
_{n}\right) -\rho _{0}\left( \mathbf{r}_{n}\right) \right) \frac{\delta
^{n}G^{\gamma }\left[ \rho _{0}\right] }{\delta \rho \left( \mathbf{r}%
_{1}\right) ...\delta \rho \left( \mathbf{r}_{n}\right) },  \label{ds1}
\end{eqnarray}%
then (\ref{v12}) implies that
\begin{equation}
0=\sum_{n=2}^{\infty }\frac{1}{n!}\int d^{3}r_{1}...d^{3}r_{n}\left( \rho
_{N}\left( \mathbf{r}_{1}\right) -\rho _{N-1}^{\gamma }\left( \mathbf{r}%
_{1}\right) \right) ...\left( \rho _{N}\left( \mathbf{r}_{n}\right) -\rho
_{N-1}^{\gamma }\left( \mathbf{r}_{n}\right) \right) \frac{\delta
^{n}G^{\gamma }\left[ \rho _{N}\right] }{\delta \rho _{N}\left( \mathbf{r}%
_{1}\right) ...\delta \rho _{N}\left( \mathbf{r}_{n}\right) },  \label{ds2}
\end{equation}%
for $F^{\gamma }[\rho ],$ $T^{\gamma }[\rho ],$ $V_{ee}^{\gamma }\left[ \rho %
\right] ,$ $T_{c}^{\gamma }\left[ \rho \right] ,$ $T^{0}\left[ \rho \right] $
and $E_{hxc}^{\gamma }\left[ \rho \right] .$ Equation (\ref{ds2}) in itself
places a constraint on the structure of potential approximate density
functionals.

Note that for $\gamma =0,$%
\begin{equation}
\rho _{N-L}^{0}\left( \mathbf{r}\right) -\rho _{N-1-L}^{0}\left( \mathbf{r}%
\right) =\left\vert \phi _{N-L}\left( \mathbf{r}\right) \right\vert ^{2}
\label{ds3}
\end{equation}%
where $\phi _{i}\left( \mathbf{r}\right) $ is an eigenfunction of the
Kohn-Sham Hamiltonian with corresponding eigenvalue $\varepsilon _{i}$
ordered so that $\varepsilon _{i}\leq \varepsilon _{i+1}.$ Equation (\ref{c4}%
), for $\gamma =0,$ becomes%
\begin{equation}
G^{\gamma }\left[ \rho _{N}\right] =\sum_{L=0}^{N-2}\int d^{3}r\left\vert
\phi _{N-L}\left( \mathbf{r}\right) \right\vert ^{2}\frac{\delta G^{\gamma }%
\left[ \sum_{i=1}^{N-L}\left\vert \phi _{i}\left( \mathbf{r}\right)
\right\vert ^{2}\right] }{\delta \rho _{N-L}^{0}\left( \mathbf{r}\right) },
\label{ds4}
\end{equation}%
and $\rho _{N-L}^{0}\left( \mathbf{r}\right) =\sum_{i=1}^{N-L}\left\vert
\phi _{i}\left( \mathbf{r}\right) \right\vert ^{2}.$ Eq. (\ref{c4}) is valid
for $G^{\gamma }\left[ \rho _{N}\right] =E_{hxc}^{\gamma }\left[ \rho _{N}%
\right] $ and for $G^{\gamma }\left[ \rho _{N}\right] =\frac{1}{\gamma }%
E_{hxc}^{\gamma }\left[ \rho _{N}\right] $ and since lim$_{\gamma
\rightarrow 0}\frac{1}{\gamma }E_{c}^{\gamma }\left[ \rho _{N}\right] =0$%
\cite{LevyPerdew:85}, it follows from (\ref{v9}) that
\begin{equation}
E_{hx}\left[ \rho _{N}\right] =\sum_{L=0}^{N-2}\int d^{3}r\left\vert \phi
_{N-L}\left( \mathbf{r}\right) \right\vert ^{2}v_{hx}\left[
\sum_{i=1}^{N-L}\left\vert \phi _{i}\left( \mathbf{r}\right) \right\vert ^{2}%
\right] .  \label{ds5}
\end{equation}%
From the definition of the Hartree potential,%
\begin{equation*}
v_{h}\left( \left[ \rho \right] ;\mathbf{r}\right) =\int d^{3}\mathbf{r}%
^{\prime }\frac{\rho \left( \mathbf{r}^{\prime }\right) }{\left\vert \mathbf{%
r-r}^{\prime }\right\vert },
\end{equation*}%
it follows form Eq. (\ref{id6}) and (\ref{v9}) that
\begin{eqnarray}
&&E_{x}\left[ \rho _{N}\right] -E_{x}\left[ \rho _{N-1}^{0}\right]  \notag \\
&=&\int d^{3}rv_{x}\left( \left[ \rho _{N}\right] ;\mathbf{r}\right)
\left\vert \phi _{N}\left( \mathbf{r}\right) \right\vert ^{2}  \notag \\
&&+\frac{1}{2}\int d^{3}rd^{3}r^{\prime }\left\vert \phi _{N}\left( \mathbf{r%
}\right) \right\vert ^{2}\frac{1}{\left\vert \mathbf{r-r}^{\prime
}\right\vert }\left\vert \phi _{N}\left( \mathbf{r}\right) \right\vert ^{2}.
\label{ds6}
\end{eqnarray}%
Eqs. (\ref{ds5}) and (\ref{ds6}) were discovered by Levy and G\"{o}rling
some time ago. \cite{LevyGorlingb:95,LevyGorlingb:95}

The only other identity that survives at $\gamma =0$ is the independent
particle Kohn-Sham kinetic energy:%
\begin{eqnarray}
&&T^{0}\left[ \rho _{N}\right] -T^{0}\left[ \rho _{N-1}^{0}\right]  \notag \\
&=&\int d^{3}r\frac{\delta T^{0}\left[ \rho _{N}\right] }{\delta \rho
_{N}\left( \mathbf{r}\right) }\left\vert \phi _{N}\left( \mathbf{r}\right)
\right\vert ^{2}.  \label{ds7}
\end{eqnarray}%
This expression follows trivially form the Kohn-Sham expression for the
independent particle kinetic energy. All identities are valid at $\gamma
\neq 0.$

In summary, equations that a set of exact density functionals satisfy for
functionals of different densities, where the densities are derived from the
same extern potential, were derived. As a corollary, it was shown that the
functionals can be expressed as a sum over integrals of functional
derivatives, where the sum runs over all $w$-representable densities for the
same external potential that integrate to an integer less than the particle
number. These relations place stringent constraints on functionals that
appear in density functional theory and it will be difficult to satisfy by
approximate functionals.

\appendix

\section{}

From Eq. (\ref{ec1})
\begin{eqnarray}
&&\frac{E_{c}^{\gamma }\left[ \rho _{N-1}^{\gamma }\right] -T_{c}^{\gamma }%
\left[ \rho _{N-1}^{\gamma }\right] }{\gamma }  \notag \\
&=&\left\langle \Psi _{\rho _{N-1}^{\gamma }}^{\gamma }\left\vert \hat{V}%
_{ee}\right\vert \Psi _{\rho _{N-1}^{\gamma }}^{\gamma }\right\rangle
-\left\langle \Psi _{\rho _{N-1}^{\gamma }}^{0}\left\vert \hat{V}%
_{ee}\right\vert \Psi _{\rho _{N-1}^{\gamma }}^{0}\right\rangle .
\label{tc1}
\end{eqnarray}%
The derivative of $E_{c}^{\gamma }\left[ \rho _{N-1}^{\gamma }\right] $ with
respect to $\gamma ,$ from definition (\ref{ec1}), can therefore be
expressed as
\begin{eqnarray}
&&\frac{\partial }{\partial \gamma }E_{c}^{\gamma }\left[ \rho
_{N-1}^{\gamma }\right]  \notag \\
&=&\frac{E_{c}^{\gamma }\left[ \rho _{N-1}^{\gamma }\right] -T_{c}^{\gamma }%
\left[ \rho _{N-1}^{\gamma }\right] }{\gamma }  \notag \\
&&+\left\langle \frac{\partial }{\partial \gamma }\Psi _{\rho _{N-1}^{\gamma
}}^{\gamma }\left\vert \hat{T}+\gamma \hat{V}_{ee}\right\vert \Psi _{\rho
_{N-1}^{\gamma }}^{\gamma }\right\rangle  \notag \\
&&+\left\langle \Psi _{\rho _{N-1}^{\gamma }}^{\gamma }\left\vert \hat{T}%
+\gamma \hat{V}_{ee}\right\vert \frac{\partial }{\partial \gamma }\Psi
_{\rho _{N-1}^{\gamma }}^{\gamma }\right\rangle  \notag \\
&&-\left\langle \frac{\partial }{\partial \gamma }\Psi _{\rho _{N-1}^{\gamma
}}^{0}\left\vert \hat{T}+\gamma \hat{V}_{ee}\right\vert \Psi _{\rho
_{N-1}^{\gamma }}^{0}\right\rangle  \notag \\
&&-\left\langle \Psi _{\rho _{N-1}^{\gamma }}^{0}\left\vert \hat{T}+\gamma
\hat{V}_{ee}\right\vert \frac{\partial }{\partial \gamma }\Psi _{\rho
_{N-1}^{\gamma }}^{0}\right\rangle .  \label{tc2}
\end{eqnarray}%
Upon adding and subtracting (c.c. stands for the complex conjugate of the
previous term)
\begin{eqnarray}
&&\left( \left\langle \frac{\partial }{\partial \gamma }\Psi _{\rho
_{N-1}^{\gamma }}^{\gamma }\left\vert \hat{v}_{N-1,\text{ext}}^{\gamma }%
\left[ \rho _{N}\right] \right\vert \Psi _{\rho _{N-1}^{\gamma }}^{\gamma
}\right\rangle +\text{c.c}\right)  \notag \\
&&+\left( \left\langle \left. \frac{\partial }{\partial \gamma }\Psi _{\rho
_{N-1}^{\gamma }}^{\gamma }\right\vert _{\gamma =0}\left\vert \hat{v}_{N-1,%
\text{ext}}^{0}\left[ \rho _{N}\right] \right\vert \Psi _{\rho
_{N-1}^{0}}^{0}\right\rangle +\text{c.c}\right)  \label{tc4}
\end{eqnarray}%
and utilizing the normalization of the wavefunctions which implies that%
\begin{equation}
\frac{\partial }{\partial \gamma }\left\langle \Psi _{\rho _{N-1}^{\gamma
}}^{\gamma }|\Psi _{\rho _{N-1}^{\gamma }}^{\gamma }\right\rangle =0,
\label{ec4}
\end{equation}%
Eq. (\ref{ec2}) becomes%
\begin{eqnarray}
&&\frac{\partial }{\partial \gamma }E_{c}^{\gamma }\left[ \rho
_{N-1}^{\gamma }\right]  \notag \\
&=&\frac{E_{c}^{\gamma }\left[ \rho _{N-1}^{\gamma }\right] -T_{c}^{\gamma }%
\left[ \rho _{N-1}^{\gamma }\right] }{\gamma }  \notag \\
&&-\left\langle \frac{\partial }{\partial \gamma }\Psi _{\rho _{N-1}^{\gamma
}}^{\gamma }\left\vert \hat{v}_{\text{ext}}^{\gamma }\left[ \rho _{N}\right]
\right\vert \Psi _{\rho _{N-1}^{\gamma }}^{\gamma }\right\rangle
-\left\langle \Psi _{\rho _{N-1}^{\gamma }}^{\gamma }\left\vert \hat{v}_{%
\text{ext}}^{\gamma }\left[ \rho _{N}\right] \right\vert \frac{\partial }{%
\partial \gamma }\Psi _{\rho _{N-1}^{\gamma }}^{\gamma }\right\rangle  \notag
\\
&&+\left\langle \frac{\partial }{\partial \gamma }\Psi _{\rho _{N-1}^{\gamma
}}^{0}\left\vert \hat{v}_{\text{ext}}^{0}\left[ \rho _{N}\right] \right\vert
\Psi _{\rho _{N-1}^{\gamma }}^{0}\right\rangle +\left\langle \Psi _{\rho
_{N-1}^{\gamma }}^{0}\left\vert \hat{v}_{\text{ext}}^{0}\left[ \rho _{N}%
\right] \right\vert \frac{\partial }{\partial \gamma }\Psi _{\rho
_{N-1}^{\gamma }}^{0}\right\rangle  \notag \\
&&-\gamma \frac{\partial }{\partial \gamma }\left\langle \Psi _{\rho
_{N-1}^{\gamma }}^{0}\left\vert \hat{V}_{ee}\right\vert \Psi _{\rho
_{N-1}^{\gamma }}^{0}\right\rangle .  \label{ec5}
\end{eqnarray}%
This equation can be simplified since
\begin{eqnarray}
&&\left\langle \frac{\partial }{\partial \gamma }\Psi _{\rho _{N-1}^{\gamma
}}^{\gamma }\left\vert \hat{v}_{\text{ext}}^{\gamma }\left[ \rho _{N}\right]
\right\vert \Psi _{\rho _{N-1}^{\gamma }}^{\gamma }\right\rangle
+\left\langle \Psi _{\rho _{N-1}^{\gamma }}^{\gamma }\left\vert \hat{v}_{%
\text{ext}}^{\gamma }\left[ \rho _{N}\right] \right\vert \frac{\partial }{%
\partial \gamma }\Psi _{\rho _{N-1}^{\gamma }}^{\gamma }\right\rangle  \notag
\\
&=&\int d^{3}r\frac{\partial \rho _{N-1}^{\gamma }\left( \mathbf{r}\right) }{%
\partial \gamma }\hat{v}_{\text{ext}}^{\gamma }\left( \left[ \rho _{N}\right]
;\mathbf{r}\right) .  \label{ap1}
\end{eqnarray}%
Using \cite{ParrYang:bk89,DreizlerGross:bk90}
\begin{eqnarray}
\left\langle \Psi _{\rho _{N-1}^{\gamma }}^{0}\left\vert \hat{V}%
_{ee}\right\vert \Psi _{\rho _{N-1}^{\gamma }}^{0}\right\rangle &=&E_{x}%
\left[ \rho _{N-1}^{\gamma }\right] +U\left[ \rho _{N-1}^{\gamma }\right]
\notag \\
&=&E_{hx}\left[ \rho _{N-1}^{\gamma }\right] ,  \label{ux1}
\end{eqnarray}%
where $E_{hx}\left[ \rho _{N-1}^{\gamma }\right] $ the sum of the exchange $%
E_{x}\left[ \rho _{N-1}^{\gamma }\right] $ and Hartree interaction energy $U%
\left[ \rho _{N-1}^{\gamma }\right] $ of the $\left( N-1\right) $-electron
system and the charge density $\rho _{N-1}^{\gamma }$ is a function of $%
\gamma $ \cite{ParrYang:bk89},%
\begin{equation}
\frac{\partial }{\partial \gamma }\left\langle \Psi _{\rho _{N-1}^{\gamma
}}^{0}\left\vert \hat{V}_{ee}\right\vert \Psi _{\rho _{N-1}^{\gamma
}}^{0}\right\rangle =\int d^{3}r\frac{\partial \rho _{N-1}^{\gamma }\left(
\mathbf{r}\right) }{\partial \gamma }v_{hx}\left( \left[ \rho _{N-1}^{\gamma
}\right] ;\mathbf{r}\right) ,  \label{ux2}
\end{equation}%
where
\begin{equation}
v_{hx}\left( \left[ \rho _{N-1}^{\gamma }\right] ;\mathbf{r}\right) =\frac{%
\delta }{\delta \rho _{N-1}^{\gamma }\left( \mathbf{r}\right) }\left( E_{x}%
\left[ \rho _{N-1}^{\gamma }\right] +U\left[ \rho _{N-1}^{\gamma }\right]
\right)  \label{ux3}
\end{equation}%
is the sum of the exchange and Hartree potentials for the $\left( N-1\right)
$-electron system. Using Eqs. (\ref{ap1}), (\ref{ux2}), (\ref{e1}) and the
fact that $\left\vert \Psi _{\rho _{N-1}^{\gamma }}^{\gamma }\right\rangle $
and $\left\vert \Psi _{\rho _{N-1}^{\gamma }}^{0}\right\rangle $ yield the
same density $\rho _{N-1}^{\gamma }$ by construction, Eq. (\ref{ec5}) can be
expressed as%
\begin{eqnarray}
&&\frac{\partial }{\partial \gamma }E_{c}^{\gamma }\left[ \rho
_{N-1}^{\gamma }\right]  \notag \\
&=&\frac{E_{c}^{\gamma }\left[ \rho _{N-1}^{\gamma }\right] -T_{c}^{\gamma }%
\left[ \rho _{N-1}^{\gamma }\right] }{\gamma }  \notag \\
&&+\int d^{3}r\frac{\partial \rho _{N-1}^{\gamma }\left( \mathbf{r}\right) }{%
\partial \gamma }\left( v_{c}^{\gamma }\left( \left[ \rho _{N}\right] ;%
\mathbf{r}\right) +\gamma v_{hx}\left( \left[ \rho _{N}\right] ;\mathbf{r}%
\right) -\gamma v_{hx}\left( \left[ \rho _{N-1}^{\gamma }\right] ;\mathbf{r}%
\right) \right)  \label{apf}
\end{eqnarray}


\begin{thebibliography}{23}
\expandafter\ifx\csname natexlab\endcsname\relax\def\natexlab#1{#1}\fi
\expandafter\ifx\csname bibnamefont\endcsname\relax
  \def\bibnamefont#1{#1}\fi
\expandafter\ifx\csname bibfnamefont\endcsname\relax
  \def\bibfnamefont#1{#1}\fi
\expandafter\ifx\csname citenamefont\endcsname\relax
  \def\citenamefont#1{#1}\fi
\expandafter\ifx\csname url\endcsname\relax
  \def\url#1{\texttt{#1}}\fi
\expandafter\ifx\csname urlprefix\endcsname\relax\def\urlprefix{URL }\fi
\providecommand{\bibinfo}[2]{#2}
\providecommand{\eprint}[2][]{\url{#2}}

\bibitem[Kohn and Sham(1965)]{KohnSham:65} \bibinfo{author}{%
\bibfnamefont{W.}~\bibnamefont{Kohn}} and \bibinfo{author}{%
\bibfnamefont{L.~J.} \bibnamefont{Sham}}, \bibinfo{journal}{Phys. Rev. A}
\textbf{\bibinfo{volume}{140}}, \bibinfo{pages}{1133} (\bibinfo{year}{1965}).

\bibitem[Hohenberg and Kohn(1964)]{HohenbergKohn:64} \bibinfo{author}{%
\bibfnamefont{P.}~\bibnamefont{Hohenberg}} and \bibinfo{author}{%
\bibfnamefont{W.}~\bibnamefont{Kohn}}, \bibinfo{journal}{Phys. Rev. B}
\textbf{\bibinfo{volume}{136}}, \bibinfo{pages}{864} (\bibinfo{year}{1964}).

\bibitem[Ruzsinszky and Perdew(2011)]{AdriennRuzsinszky2011} %
\bibinfo{author}{\bibfnamefont{A.}~\bibnamefont{Ruzsinszky}} and %
\bibinfo{author}{\bibfnamefont{J.~P.} \bibnamefont{Perdew}}, %
\bibinfo{journal}{Computational and Theoretical Chemistry} \textbf{%
\bibinfo{volume}{963}}, \bibinfo{pages}{2} (\bibinfo{year}{2011}).

\bibitem[Perdew et~al.(2010)Perdew, Ruzsinszky, Csonka, Constantin, and Sun]%
{JohnP.Perdew2009}
\bibinfo{author}{\bibfnamefont{J.~P.}
\bibnamefont{Perdew}}, \bibinfo{author}{\bibfnamefont{A.}~%
\bibnamefont{Ruzsinszky}},
\bibinfo{author}{\bibfnamefont{G.~I.}
\bibnamefont{Csonka}},
\bibinfo{author}{\bibfnamefont{L.~A.}
\bibnamefont{Constantin}}, and \bibinfo{author}{\bibfnamefont{J.}~%
\bibnamefont{Sun}}, \bibinfo{journal}{Phys. Rev. Lett.} \textbf{%
\bibinfo{volume}{103}}, \bibinfo{pages}{026403} (\bibinfo{year}{2010}).

\bibitem[Csonka et~al.(2009)Csonka, Perdew, Ruzsinszky, Philipsen,
Leb\`egue, Paier, Vydrov, and \'Angy\'an]{GaborI.Csonka2009a} %
\bibinfo{author}{\bibfnamefont{G.~I.} \bibnamefont{Csonka}}, %
\bibinfo{author}{\bibfnamefont{J.~P.} \bibnamefont{Perdew}}, %
\bibinfo{author}{\bibfnamefont{A.}~\bibnamefont{Ruzsinszky}}, %
\bibinfo{author}{\bibfnamefont{P.~H.~T.} \bibnamefont{Philipsen}}, %
\bibinfo{author}{\bibfnamefont{S.}~\bibnamefont{Leb\`egue}}, %
\bibinfo{author}{\bibfnamefont{J.}~\bibnamefont{Paier}}, \bibinfo{author}{%
\bibfnamefont{O.~A.} \bibnamefont{Vydrov}}, and
\bibinfo{author}{\bibfnamefont{J.~G.}
  \bibnamefont{\'Angy\'an}}, \bibinfo{journal}{Phys. Rev. B} \textbf{%
\bibinfo{volume}{79}}, \bibinfo{pages}{155107} (\bibinfo{year}{2009}).

\bibitem[Staroverov et~al.(2004)Staroverov, Scuseria, Tao, and Perdew]%
{Staroverov2004}
\bibinfo{author}{\bibfnamefont{V.~N.}
\bibnamefont{Staroverov}},
\bibinfo{author}{\bibfnamefont{G.~E.}
\bibnamefont{Scuseria}}, \bibinfo{author}{\bibfnamefont{J.}~%
\bibnamefont{Tao}}, and
\bibinfo{author}{\bibfnamefont{J.~P.}
\bibnamefont{Perdew}}, \bibinfo{journal}{Phys. Rev. B} \textbf{%
\bibinfo{volume}{69}}, \bibinfo{pages}{075102} (\bibinfo{year}{2004}).

\bibitem[Tao et~al.(2003)Tao, Perdew, Staroverov, and Scuseria]{TaoPerdew03} %
\bibinfo{author}{\bibfnamefont{J.}~\bibnamefont{Tao}}, \bibinfo{author}{%
\bibfnamefont{J.~P.} \bibnamefont{Perdew}}, \bibinfo{author}{%
\bibfnamefont{V.~N.} \bibnamefont{Staroverov}}, and
\bibinfo{author}{\bibfnamefont{G.~E.}
  \bibnamefont{Scuseria}}, \bibinfo{journal}{Phys. Rev. Lett.} \textbf{%
\bibinfo{volume}{91}}, \bibinfo{pages}{146401} (\bibinfo{year}{2003}).

\bibitem[Perdew et~al.(2005)Perdew, Ruzsinszky, and Tao]{JohnP.Perdew2005} %
\bibinfo{author}{\bibfnamefont{J.~P.} \bibnamefont{Perdew}}, %
\bibinfo{author}{\bibfnamefont{A.}~\bibnamefont{Ruzsinszky}}, and %
\bibinfo{author}{\bibfnamefont{J.}~\bibnamefont{Tao}},
\bibinfo{journal}{J.
Chem. Phys.} \textbf{\bibinfo{volume}{123}}, \bibinfo{pages}{062201} (%
\bibinfo{year}{2005}).

\bibitem[Harris and Jones(1974)]{HarrisJones:74} \bibinfo{author}{%
\bibfnamefont{J.}~\bibnamefont{Harris}} and \bibinfo{author}{%
\bibfnamefont{R.~O.} \bibnamefont{Jones}}, \bibinfo{journal}{J. Phys. F}
\textbf{\bibinfo{volume}{4}}, \bibinfo{pages}{1174} (\bibinfo{year}{1974}).

\bibitem[Langreth and Perdew(1975)]{LangrethPerdew:75} \bibinfo{author}{%
\bibfnamefont{D.~C.} \bibnamefont{Langreth}} and \bibinfo{author}{%
\bibfnamefont{J.~P.} \bibnamefont{Perdew}},
\bibinfo{journal}{Solid State
Comm.} \textbf{\bibinfo{volume}{17}}, \bibinfo{pages}{1425} (%
\bibinfo{year}{1975}).

\bibitem[Langreth and Perdew(1977)]{LangrethPerdew:77} \bibinfo{author}{%
\bibfnamefont{D.~C.} \bibnamefont{Langreth}} and \bibinfo{author}{%
\bibfnamefont{J.~P.} \bibnamefont{Perdew}}, \bibinfo{journal}{Phys. Rev. B}
\textbf{\bibinfo{volume}{15}}, \bibinfo{pages}{2884} (\bibinfo{year}{1977}).

\bibitem[Gunnarson and Lundqvist(1976)]{GunnarsonLundqvist:76} %
\bibinfo{author}{\bibfnamefont{O.}~\bibnamefont{Gunnarson}} and %
\bibinfo{author}{\bibfnamefont{B.~I.} \bibnamefont{Lundqvist}}, %
\bibinfo{journal}{Phys. Rev. B} \textbf{\bibinfo{volume}{13}}, %
\bibinfo{pages}{4274} (\bibinfo{year}{1976}).

\bibitem[Levy(1979)]{Levy:79} \bibinfo{author}{\bibfnamefont{M.}~%
\bibnamefont{Levy}}, \bibinfo{journal}{Natl.
  Acad. Sci. USA} \textbf{\bibinfo{volume}{76}}, \bibinfo{pages}{6062} (%
\bibinfo{year}{1979}).

\bibitem[Levy and Perdew(1985)]{LevyPerdew:85} \bibinfo{author}{%
\bibfnamefont{M.}~\bibnamefont{Levy}} and \bibinfo{author}{%
\bibfnamefont{J.~P.} \bibnamefont{Perdew}}, \bibinfo{journal}{Phys. Rev. A}
\textbf{\bibinfo{volume}{32}}, \bibinfo{pages}{2010} (\bibinfo{year}{1985}).

\bibitem[G{\"o}rling and Levy(1993)]{GorlingLevy:93} \bibinfo{author}{%
\bibfnamefont{A.}~\bibnamefont{G{\"o}rling}} and \bibinfo{author}{%
\bibfnamefont{M.}~\bibnamefont{Levy}}, \bibinfo{journal}{Phys. Rev. B}
\textbf{\bibinfo{volume}{47}}, \bibinfo{pages}{13105} (\bibinfo{year}{1993}).

\bibitem[Parr and Yang(1989)]{ParrYang:bk89} \bibinfo{author}{%
\bibfnamefont{R.~G.} \bibnamefont{Parr}} and \bibinfo{author}{%
\bibfnamefont{W.}~\bibnamefont{Yang}}, \emph{%
\bibinfo{title}{Density
Functional Theory of Atoms and Molecules}} (%
\bibinfo{publisher}{Oxford
University Press}, \bibinfo{address}{New York}, \bibinfo{year}{1989}).

\bibitem[Dreizler and Gross(1990)]{DreizlerGross:bk90} \bibinfo{author}{%
\bibfnamefont{R.~M.} \bibnamefont{Dreizler}} and \bibinfo{author}{%
\bibfnamefont{E.~K.~U.} \bibnamefont{Gross}}, \emph{%
\bibinfo{title}{Density
Functional Theory}} (\bibinfo{publisher}{Springer-Verlag}, %
\bibinfo{address}{Berlin}, \bibinfo{year}{1990}).

\bibitem[Jones and Gunnarsson(1989)]{JonesGunnarsson:89} \bibinfo{author}{%
\bibfnamefont{R.}~\bibnamefont{Jones}} and \bibinfo{author}{%
\bibfnamefont{O.}~\bibnamefont{Gunnarsson}},
\bibinfo{journal}{Rew. Mod.
Phys.} \textbf{\bibinfo{volume}{61}}, \bibinfo{pages}{689} (%
\bibinfo{year}{1989}).

\bibitem[Levy and G{\"o}rling(1996{a})]{LevyGorling:96} \bibinfo{author}{%
\bibfnamefont{M.}~\bibnamefont{Levy}} and \bibinfo{author}{%
\bibfnamefont{A.}~\bibnamefont{G{\"o}rling}},
\bibinfo{journal}{Phys. Rev.
A} \textbf{\bibinfo{volume}{53}}, \bibinfo{pages}{3140} (%
\bibinfo{year}{1996}{\natexlab{a}}).

\bibitem[Levy and G{\"o}rling(1996{b})]{LevyGorlingb:96} \bibinfo{author}{%
\bibfnamefont{M.}~\bibnamefont{Levy}} and \bibinfo{author}{%
\bibfnamefont{A.}~\bibnamefont{G{\"o}rling}},
\bibinfo{journal}{Phys. Rev.
B} \textbf{\bibinfo{volume}{53}}, \bibinfo{pages}{969} (%
\bibinfo{year}{1996}{\natexlab{b}}).

\bibitem[Perdew et~al.(1982)Perdew, Parr, Levy, and Balduz]{PPLB:82} %
\bibinfo{author}{\bibfnamefont{J.}~\bibnamefont{Perdew}}, %
\bibinfo{author}{\bibfnamefont{R.}~\bibnamefont{Parr}}, \bibinfo{author}{%
\bibfnamefont{M.}~\bibnamefont{Levy}}, and \bibinfo{author}{%
\bibfnamefont{J.}~\bibnamefont{Balduz}}, \bibinfo{journal}{Phys. Rev. Lett.}
\textbf{\bibinfo{volume}{49}}, \bibinfo{pages}{1691} (\bibinfo{year}{1982}).

\bibitem[Joubert(2011)]{Joubert2011a}
\bibinfo{author}{\bibfnamefont{D.~P.}
\bibnamefont{Joubert}},
\bibinfo{journal}{arXiv:1107.3219v1
[cond-mat.mtrl-sci]} (\bibinfo{year}{2011}).

\bibitem[Levy and G{\"o}rling(1995)]{LevyGorlingb:95} \bibinfo{author}{%
\bibfnamefont{M.}~\bibnamefont{Levy}} and \bibinfo{author}{%
\bibfnamefont{A.}~\bibnamefont{G{\"o}rling}},
\bibinfo{journal}{Int. J.
Quantum Chem.} \textbf{\bibinfo{volume}{56}}, \bibinfo{pages}{385} (%
\bibinfo{year}{1995}).
\end{thebibliography}

\end{document}